\pdfoutput=1

\documentclass[11pt,twoside,a4paper,cmspaper,final,collab]{cms-tdr}
\def\svnVersion{63423}\def\cmsCernNo{2011-045}\def\cmsCernDate{\today}\def\cmsMessage{Submitted to Physics Letters B}

\begin{document}\cmsNoteHeader{EWK-10-008}

\hyphenation{had-ron-i-za-tion}
\hyphenation{cal-or-i-me-ter}
\hyphenation{de-vices}
\RCS$Revision: 62960 $
\RCS$HeadURL: svn+ssh://alverson@svn.cern.ch/reps/tdr2/papers/EWK-10-008/trunk/EWK-10-008.tex $
\RCS$Id: EWK-10-008.tex 62960 2011-06-21 20:59:29Z vuko $

\newcommand{\Wg}{\ensuremath{\text{W} \gamma}}%
\newcommand{\Zg}{\ensuremath{\text{Z} \gamma}}%

\newcommand{\ETG}{\ensuremath{E_T^\gamma}}
\newcommand{\totlumi}{\ensuremath{36 \pbinv}}%

\newcommand{\ecaliso}{\ensuremath{{\rm Iso}_\mathrm{ECAL}}\xspace}
\newcommand{\hcaliso}{\ensuremath{{\rm Iso}_\mathrm{HCAL}}\xspace}
\newcommand{\trkiso}{\ensuremath{{\rm Iso}_\mathrm{TRK}}\xspace}
\newcommand{\Iso}{\ensuremath{{\rm Iso}}\xspace}
\newcommand{\see}{\ensuremath{\sigma_{\eta\eta}}\xspace}
\newcommand{\sieie}{\ensuremath{\sigma_{i\eta i\eta}}\xspace}

\newcommand{\nsig}{\ensuremath{N_{sig}}}
\newcommand{\nback}{\ensuremath{N_{backg}}}
\newcommand{\nobs}{\ensuremath{N_{obs}}}

\cmsNoteHeader{EWK-10-008} 
\title{Measurement of W$\gamma$ and Z$\gamma$ production in pp collisions at $\sqrt{s} = 7$~TeV}

\date{\today}
\abstract{

A measurement of \Wg\ and \Zg\ production in proton-proton
collisions at $\sqrt{s} = 7$~TeV is presented. Results are
based on a data sample recorded by the CMS experiment at 
the LHC, corresponding to an integrated
luminosity of 36 pb$^{-1}$. The electron and muon decay channels of the W
and \Z\ are used.  The total cross sections are measured for photon
transverse energy $E_T^\gamma>10$~GeV and spatial separation from charged
leptons in the plane of pseudorapidity and azimuthal angle
$\DR(\ell,\gamma)>0.7$, and with an additional dilepton invariant mass
requirement of $M_{\ell\ell} > 50$~GeV for the \Zg\ process.
The following cross section times branching fraction values are found:
$\sigma(\mathrm{pp}\to\Wg+X) \times \mathcal{B}(\text{W} \to \ell\nu) =
56.3 \pm 5.0~\text{(stat.)} \pm 5.0~\text{(syst.)} \pm 2.3~\text{~(lumi.)}$~pb and
$\sigma(\mathrm{pp}\to\Zg+X) \times \mathcal{B}(\text{Z}\to \ell\ell) =
9.4 \pm 1.0 \text{~(stat.)} \pm 0.6 \text{~(syst.)} \pm 0.4 \text{~(lumi.)}$~pb.
These measurements are in agreement with standard model predictions.
The first limits on anomalous WW$\gamma$, ZZ$\gamma$, and
Z$\gamma\gamma$ trilinear gauge couplings at $\sqrt{s} = 7$~TeV are set.
}

\hypersetup{%
pdfauthor={CMS Collaboration},%
pdftitle={Measurement of W-gamma and Z-gamma production in pp collisions at sqrt(s) = 7 TeV},%
pdfsubject={CMS},%
pdfkeywords={CMS, physics, electroweak}}

\maketitle 

The study of Z$\gamma$ and W$\gamma$ production in proton-proton collisions
is an important test of the standard model (SM) because of its sensitivity to
the self-interaction between gauge bosons via trilinear gauge
boson couplings (TGCs). These self-interactions are a direct consequence
of the non-Abelian $SU(2) \times U(1)$ gauge symmetry of the SM and are
a necessary ingredient to construct renormalizable theories
involving massive gauge bosons that satisfy unitarity. The values of these
couplings are fully fixed in the SM by the gauge structure of the Lagrangian.
Thus, any deviation of the observed strength of the TGC from the SM
prediction would indicate new physics, for example, the production of new particles
that decay to Z$\gamma$ or W$\gamma$, or new interactions that increase the
strength of the TGCs. Previous searches for anomalous TGCs (aTGCs)
performed at lower energies by the e$^+$e$^-$ LEP~\cite{lep,l3_1,l3_2,opal,
Schael:2004tq,Schael:2009zz,delphi:2007pq,delphi:2010zj}
and $\mathrm{p\bar{p}}$ Tevatron experiments~\cite{tevatron_wg1, tevatron_wg2, tevatron_wg3, tevatron_zg0, tevatron_zg1, tevatron_zg2}
yielded results consistent with the SM. Testing TGCs at the Large Hadron
Collider (LHC) is particularly interesting because  it extends the test of the validity
of the SM  description of interactions in the bosonic sector to
substantially higher energies.

We present the first measurement of the W$\gamma$ and Z$\gamma$
cross sections, and of the WW$\gamma$, ZZ$\gamma$, and
Z$\gamma\gamma$ TGCs at $\sqrt{s} = $~7~TeV, using data collected with the
Compact Muon Solenoid (CMS) detector in 2010, corresponding to an integrated
luminosity of \totlumi.

Final-state particles in the studied collision events are reconstructed
in the CMS detector, which consists of several subdetectors.
The central tracking system is based on silicon pixel
and strip detectors, which allow the trajectories of  charged particles
to be reconstructed in the pseudorapidity
range $|\eta| < 2.5$, where $\eta = -\ln \tan(\theta/2)$
and $\theta$ is the polar angle relative to the counterclockwise proton
beam direction. CMS uses a right-handed coordinate system, in which
the $x$ axis lies in the accelerator plane and points towards
the center of the LHC ring, the $y$ axis is directed
upwards, and the $z$ axis runs along the beam axis.
Electromagnetic (ECAL) and hadron (HCAL)
calorimeters are located outside the tracking system and provide
coverage for $|\eta| < 3$.
The ECAL and HCAL are finely segmented with granularities
$\Delta\eta \times \Delta\phi = 0.0175 \times 0.0175$
and $0.087 \times 0.087$, respectively,
at central pseudorapidities and with a coarser granularity at
forward pseudorapidities; $\phi$ denotes the azimuthal angle,
measured in radians. A preshower detector
made of silicon sensor planes and lead absorbers is located in front of
the ECAL at $1.653 < |\eta| < 2.6$. The calorimeters and tracking systems
are located within the 3.8 T magnetic field of the superconducting solenoid.
Muons are measured in gas-ionization detectors embedded in the steel return
yoke. In addition to the barrel and endcap detectors, CMS includes extensive
calorimetry in the forward regions. A detailed description of CMS can be
found elsewhere~\cite{CMSdetector}.

The \Wg\  and \Zg\ processes are studied in the final states
$\ell\nu\gamma$ and $\ell\ell\gamma$, respectively, where $\ell$ is either
an electron
or a muon. Leading order (LO) W$\gamma$ production can be described by three
processes: initial state radiation (ISR), where a photon is radiated by
one of the incoming quarks; final state radiation (FSR), where a photon is radiated
from the charged lepton from the W boson decay; and finally through the
WW$\gamma$ vertex, where a photon couples directly to the W boson.
In the SM, LO \Zg\ production is described via ISR and FSR
processes only, because the ZZ$\gamma$ and Z$\gamma\gamma$
TGCs are not allowed at the tree level.

As at LO the \Wg\ and \Zg\ cross sections diverge for soft
photons or, in the case of Z/$\gamma^*\gamma$ production, for small
values of the dilepton invariant mass, we restrict the cross section measurement
to the phase space defined by the following two kinematic requirements:
the photon candidate must have transverse energy \ETG\ larger than 10~GeV,
and it must be spatially separated from the final-state charged lepton(s)
by $\Delta R(\ell, \gamma)>0.7$, where
$\DR = \sqrt{(\eta_\ell-\eta_\gamma)^{2}+(\phi_\ell-\phi_\gamma)^{2}}$.
 Furthermore, for the Z$\gamma$ final state,
the invariant mass of the two lepton candidates must be above 50~GeV.

The main background to \Wg\ and \Zg\ production consists of
W+jets and Z+jets events, respectively, where the photon candidate
originates from one of the jets. We estimate this background from data.
The contribution from other processes, such as t$\bar{\text{t}}$ and multijet
QCD production, is much smaller and it is estimated from Monte Carlo (MC)
simulation studies. All signal samples for W$\gamma+n$ jets and
Z$\gamma+n$ jets ($n \le 1$) are generated with \SHERPA~\cite{sherpa}
and further interfaced with \PYTHIA~\cite{Pythia} for showering and
hadronization.
The kinematic distributions for these signal processes are further
cross-checked with simulated samples generated
with \MADGRAPH~\cite{MadGraph} interfaced with \PYTHIA and good
agreement is found. The signal samples are normalized using the
next-to-leading order (NLO) prediction
from the NLO {\sc Baur} generator~\cite{baur}. Background processes have been
generated with the \MADGRAPH+~\PYTHIA combination for
t$\bar{\text{t}}$, W+jets, and Z+jets. Multijet QCD, $\gamma$+jets and
diboson processes are produced using only the
\PYTHIA generator. All generated samples are passed through a detailed
simulation of the CMS detector based on \GEANTfour~\cite{geant4} and the
same complete reconstruction chain used for data analysis.
All background samples are normalized to the integrated luminosity of the
data sample using NLO cross section predictions, except inclusive
W and Z production, for which the next-to-next-to-leading order cross section
is used~\cite{NNLO}.

Photon candidates are reconstructed from clusters of energy deposits
in the ECAL. We require photon candidates to be in $|\eta| < 1.44$ or
$1.57 < |\eta| < 2.5$. Photons that undergo conversion
in the material in front of the ECAL are also efficiently reconstructed
by the same clustering algorithm. The clustered energy is corrected, taking into
account interactions in the material in front of the ECAL and electromagnetic
shower containment~\cite{EGM-10-005}. The photon candidate's pseudorapidity
is calculated using the position of the primary interaction vertex.
The absolute photon energy scale is determined using electrons from
reconstructed Z boson decays with an uncertainty estimated to be less
than 2\%, and further verified using an independent FSR Z$\to \mu\mu\gamma$ 
data sample, selected with similar selection criteria used to select Z$\gamma$
candidates events but with $\Delta R(\gamma, \mu) < 0.7$,
by comparing the $\mu\mu\gamma$ invariant mass to the nominal Z boson mass.
Both the position and the width of the peak of the $\mu\mu\gamma$ invariant
mass distribution in MC simulation are found to be consistent with that observed in data.
We estimate the systematic uncertainty due to modeling of the photon energy measurement
by varying the photon energy scale and resolution in the MC simulation
within the uncertainties of the data-MC simulation agreement of the $\mu\mu\gamma$
invariant mass distribution.
To reduce the background from electrons, photon candidates must not have
associated hits in the innermost layer of the pixel subdetector.
To reduce the background from misidentified jets, photon clusters are
required to be isolated from other activity in the ECAL, HCAL, and
tracker system. This photon isolation is defined by requiring 
the scalar sum of transverse energies or momenta reconstructed
in the HCAL, ECAL, and Tracker sub-detectors, and spatially separated
from the photon candidate by $\Delta R < 0.4$, to be less
than 4.2, 2.2, and 2.0 \gev, respectively.
Finally, the photon candidate's energy deposition profile in pseudorapidity
must be consistent with the shape expected for a photon~\cite{EGM-10-005}.
The adopted photon selection criteria lead to a signal efficiency of about 90\%,
while significantly suppressing the major background from misidentified jets.

Electron candidates are reconstructed from clusters of energy deposited in
the ECAL that are matched to a charged track reconstructed in the silicon
tracker. Similar requirements to those for photon candidates are applied
to the ECAL energy cluster. We require electron candidates to
have $\pt>20$~GeV and $|\eta|<2.5$.
Two sets of electron identification criteria based on
shower shape and track-cluster spatial matching are applied to the reconstructed
candidates. These criteria are designed to reject misidentified jets from QCD multijet
production while maintaining at least 80\% (95\%) efficiency for electrons from the
decay of W or Z bosons for the tighter (looser) criteria.  This efficiency is
defined relative to the sample of reconstructed electrons.
The tighter set of criteria is the same as the one used in the CMS measurement of
the W and Z boson cross sections~\cite{wzxs}.
Electrons originating from photon conversions are suppressed by dedicated
algorithms~\cite{electronID}.
The tighter selection is used for the W$\gamma$ final state, while the looser
selection is used for Z$\gamma$.

Muons are reconstructed as charged tracks matched to hits and segments
in the muon system. The track associated with the muon candidate is required to
have at least 11 hits in the silicon tracker, it must be consistent
with originating from the primary vertex in the event, and it must
be spatially well-matched to the muon system including a minimum number
of hits in the muon detectors. These selection criteria follow the standard
muon identification requirements employed in previous analyses~\cite{wzxs}
that are 95\% efficient for muons produced in W and Z boson decays.
All muon candidates are required to have $\pt>20\GeV$
and $|\eta|< 2.4$. The muon candidates in W$\gamma \to \mu\nu\gamma$
are further restricted to be in the fiducial volume of the single muon trigger,
$|\eta| < 2.1$.

All lepton identification and reconstruction efficiencies of final state
particles are measured in data using Z$\to \ell^+\ell^-$ events~\cite{wzxs}
and are found to be within a few percent of those obtained from MC simulation.

\begin{figure}[tb]
  \begin{center}
    \includegraphics[width=0.49\textwidth]{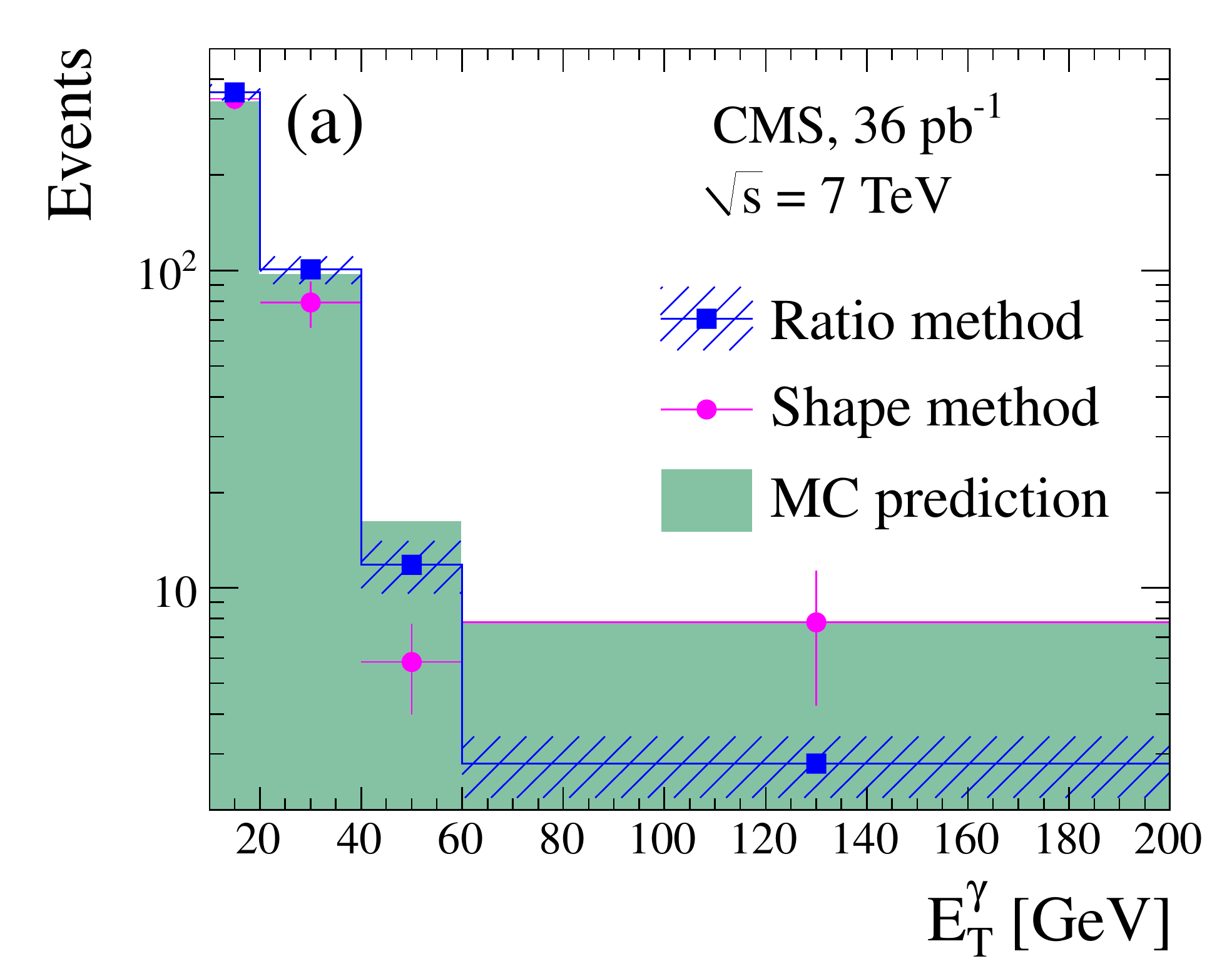}
    \includegraphics[width=0.49\textwidth]{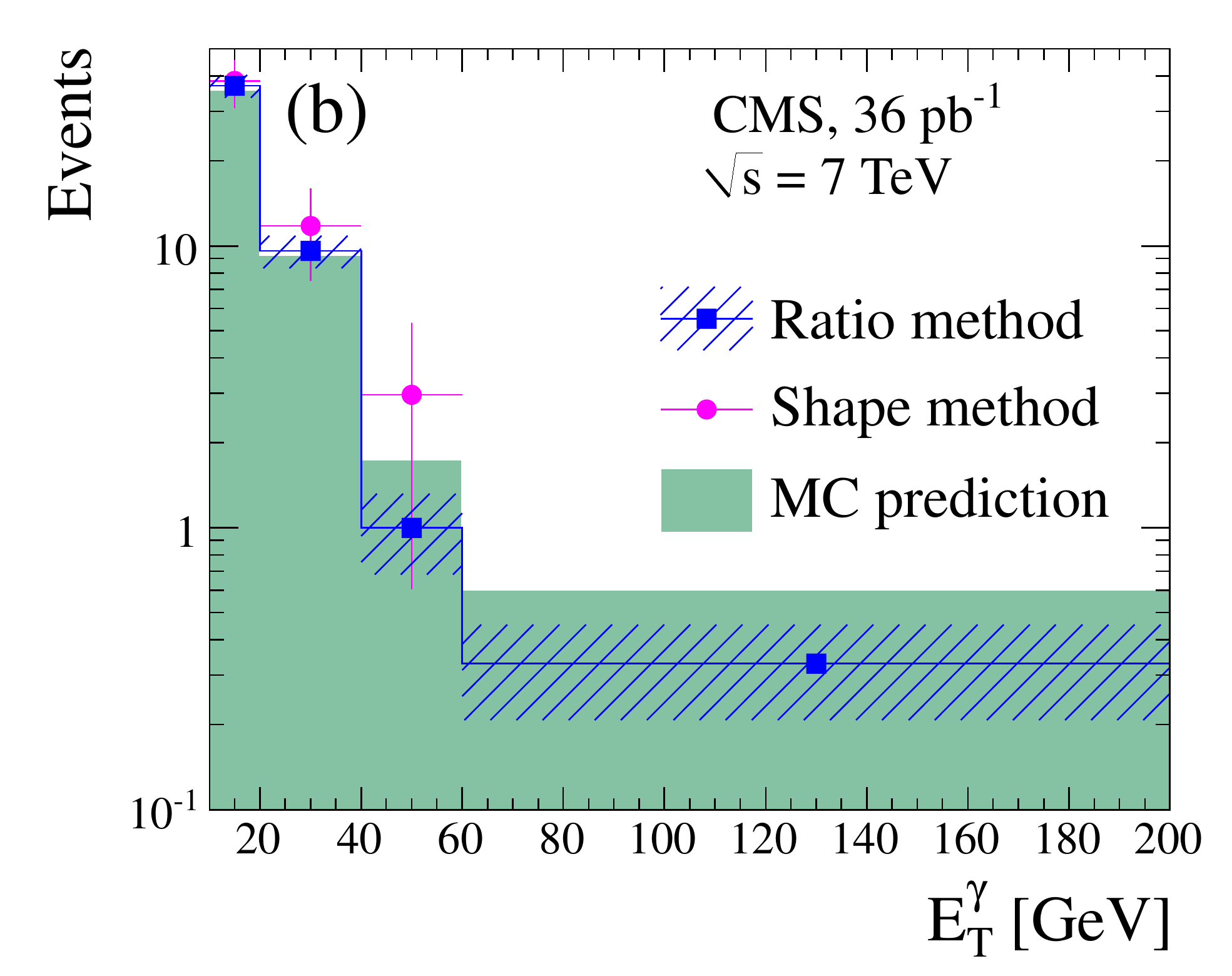}
    \caption{
Background from misidentified jets as a function of the photon candidate $E_\text{T}$,
estimated from the ratio method, is shown with blue squares together with
an alternative method that uses energy deposition shape templates
(magenta circles), and MC simulation (green filled histogram) for (a)  \Wg\
and (b) \Zg\ channels. Uncertainties include both statistical and systematic
sources.
}
\label{fig:backgrounds}
 \end{center}
\end{figure}

To estimate the background due to jets misidentified as photons, we
use a method based on the assumption that the properties of
jets misidentified as photons do not depend on the jet production mechanism and that photon
candidates originating in jets in W+jets and Z+jets events are similar
to those in multijet QCD events. We estimate the W+jets and Z+jets
background contributions by measuring the $E_\text{T}$-dependent
probability for a jet to be identified as a photon candidate,
and then folding this probability with the nonisolated photon
candidate $E_\text{T}$ spectrum observed in the \Wg\ and \Zg\ samples.
The former is measured in a sample of multijet QCD events containing at least one
high-quality jet candidate that satisfies the CMS jet trigger
requirement~\cite{cmsjets}. Any photon candidate observed in such a
sample is most likely a misidentified jet. We then measure
the \ETG-dependent ratio of jets passing the full photon
identification criteria to those identified as
photons but failing the track isolation requirement.
As the contribution from genuine photons in the multijet sample
from $\gamma$+jets processes becomes significant at large values
of \ETG, we subtract this contribution from the
total ratio using a Monte Carlo simulation prediction. The obtained
$E_\text{T}$-dependent probability is folded with the nonisolated photon
candidates in the W$\gamma$ and Z$\gamma$ candidate events to estimate the
number of W+jets and Z+jets events, respectively, passing the full selection criteria.
The estimation of the background from misidentified jets
for the W$\gamma$ and Z$\gamma$ processes is further cross-checked
with W+jets and Z+jets MC simulation and with the results obtained from
an independent study of photon cluster shower shapes
following the same approach as in Ref.~\cite{Khachatryan:2010fm} (shape
method).
We observe good agreement between all three methods (Fig.~\ref{fig:backgrounds}).

A neutrino from leptonic W boson decay does not interact with the
detector and results in a significant missing transverse energy, \MET,
in the event. The \MET in this analysis is calculated with the
particle-flow method~\cite{pf}. The algorithm combines information from
the tracking system, the muon chambers, and from all the calorimetry to classify
reconstructed objects according to their particle type (electron, muon,
photon, charged or neutral hadron). This allows precise corrections
to particle energies and also provides a significant degree of redundancy,
which renders the \MET measurement less sensitive to calorimetry miscalibration.
The \MET is computed as the magnitude of the negative vector sum of transverse
energies of all particle-flow objects. Both ECAL and HCAL are known
to record anomalous signals that correspond to particles hitting the
transducers, or to rare random discharges of the readout detectors.
Anomalous noise in the calorimeters can reduce the accuracy of the \MET measurement.
Algorithms designed to suppress such noise reduce it to a negligible
level, as shown in studies based on cosmic rays and control samples~\cite{anomalousNoiseMET}.
The modeling of \MET in the simulation is checked using events with
(W$\to\ell\nu$) and without (Z$\to\ell^+\ell^-$) genuine \MET and good agreement
is found~\cite{met,wzxs}.

\begin{figure}[tb]
  \begin{center}
    \includegraphics[width=0.8\textwidth]{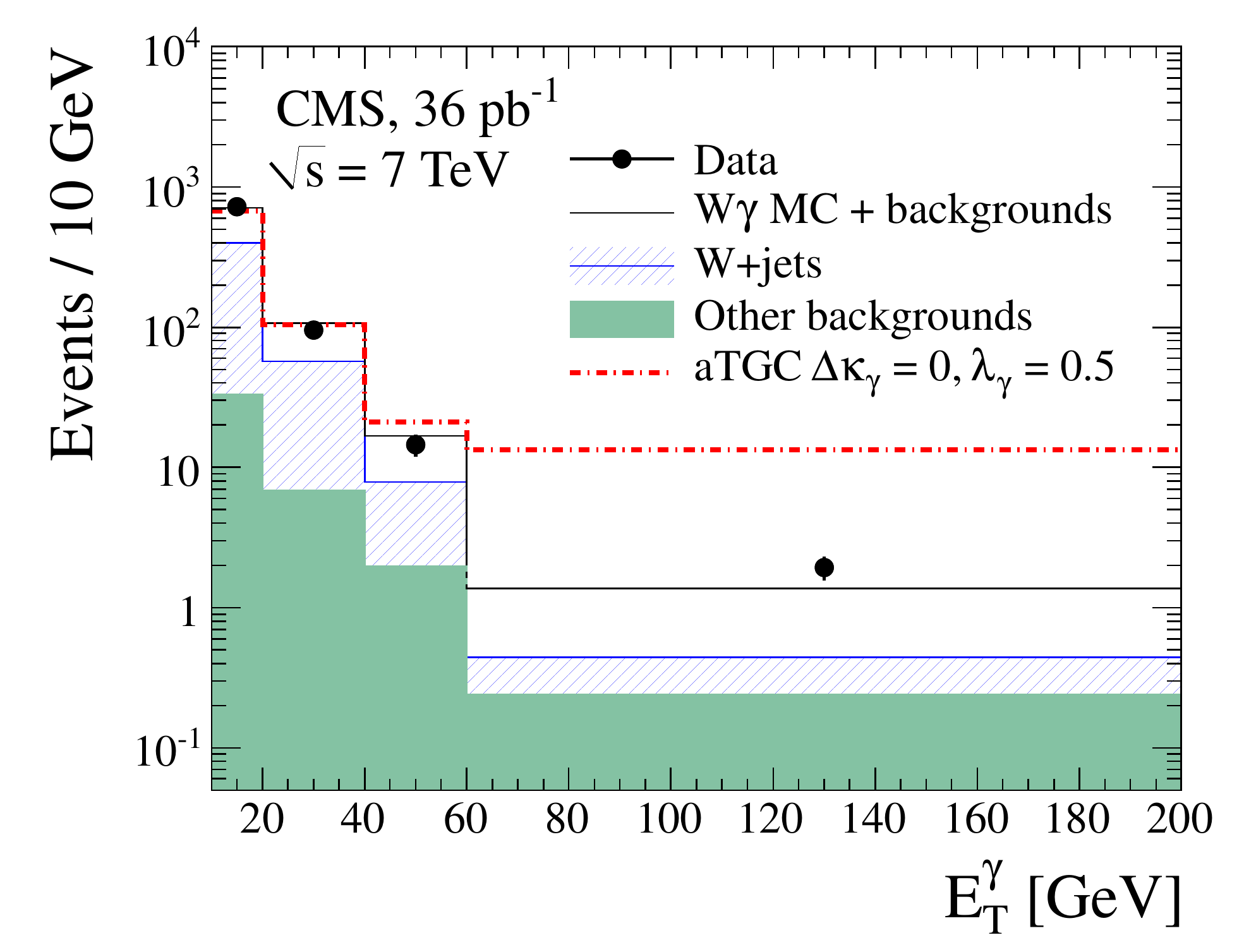}
    \caption{
Transverse energy distribution for the photon candidates for W$\gamma$
production. Data are shown with black circles with error bars; expected
signal plus background is shown as a black solid histogram; the contribution from
misidentified jets is given as a hatched blue histogram, and the background
from $\gamma+\text{jets}$, $\mathrm{t\bar{t}}$, and multiboson processes is given as
a solid green histogram. A typical aTGC signal is given as a red
dot-and-line histogram. The last bin includes overflows. 
Entries in wider bins are normalized to the ratio of 10 GeV and the bin width.}
   \label{fig:ETG_wg}
  \end{center}
\end{figure}

\begin{figure}[p!]
  \begin{center}
    \includegraphics[width=0.95\textwidth]{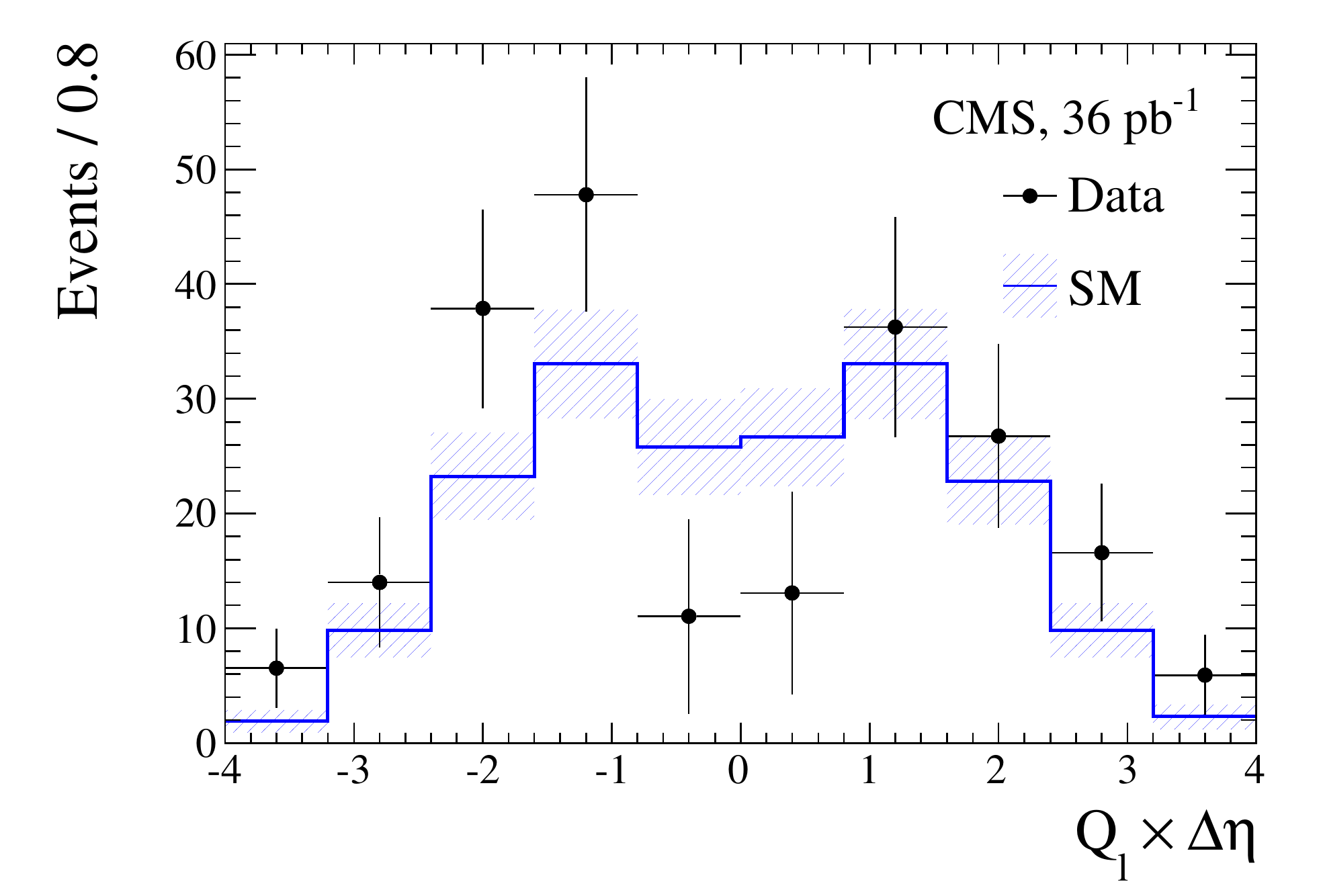}
    \caption{The background-subtracted charge-signed rapidity
    difference for the combined electron and muon channels of
    W$\gamma$ production is shown for data (black circles with error
    bars) and SM simulation (blue hatched region).
    The results of the Kolmogorov-Smirnov test of the agreement
    between data and MC prediction is 57\%, which indicates a
    reasonable agreement.
    }
    \label{fig:raz}
  \end{center}
\end{figure}

\begin{figure}[tb]
  \begin{center}
    \includegraphics[width=0.8\textwidth]{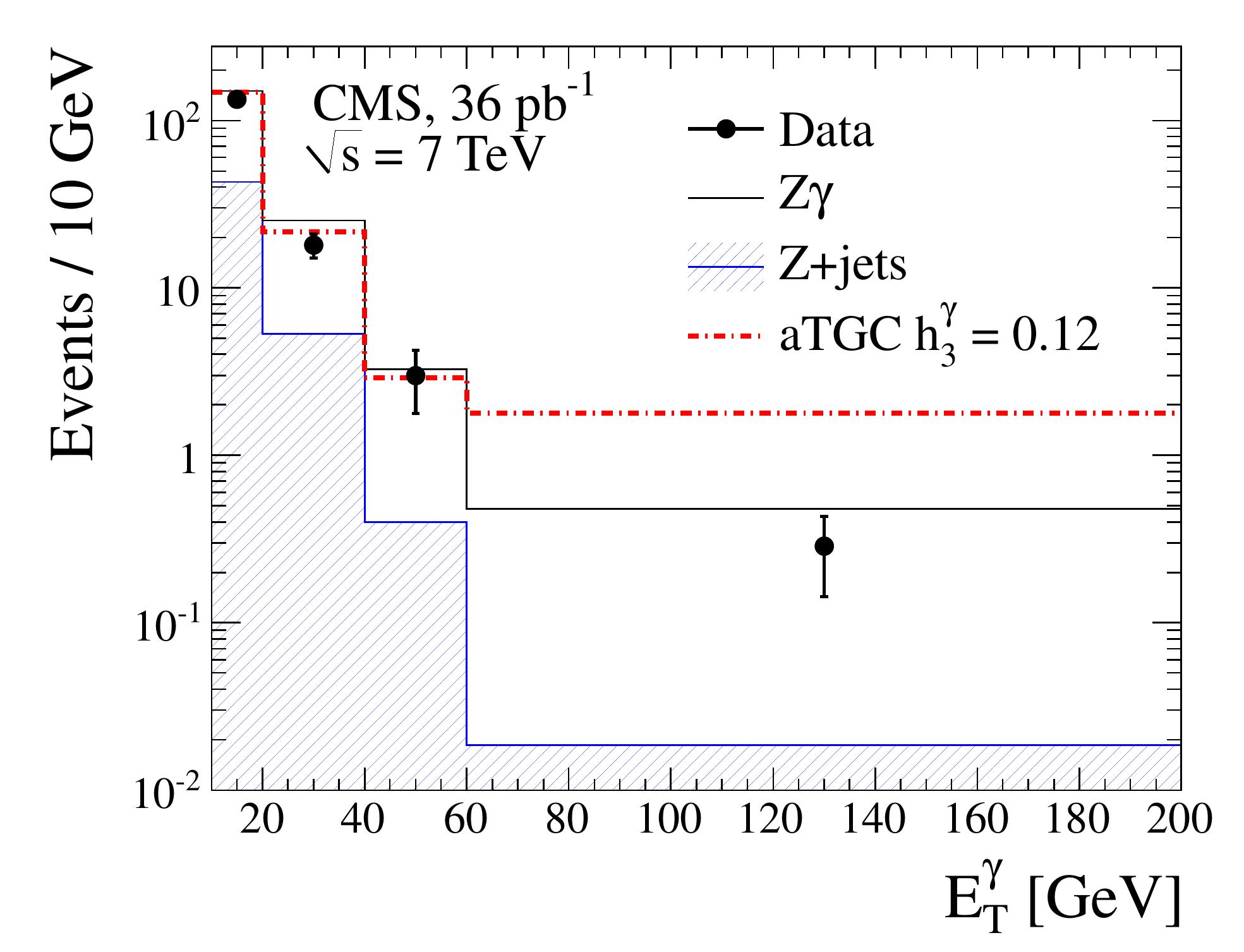}
    \caption{
     The transverse energy distribution of photon candidates in the \Zg\ channel
     in data is shown with black circles with error bars;
     the expected signal plus background is shown
     as a solid black histogram, while the contribution from misidentified jets is given as a
     hatched blue histogram. A typical aTGC signal is given as a red
     dot-and-line histogram. The last bin includes overflows.
     Entries in wider bins are normalized to the ratio of 10 GeV and the bin width.}
   \label{fig:ETG_zg}
  \end{center}
\end{figure}

Data for this study are selected with the CMS two-level trigger system by
requiring the events to have at least one energetic electron or muon,
consistent with being produced from W or Z boson decays.
This requirement is about 90\% efficient for the W$\gamma\to\mu\nu \gamma$
signal and 98\% efficient for W$\gamma\to \text{e}\nu \gamma$. The trigger
efficiency is close to 100\% for both Z$\gamma\to \ell\ell\gamma$ final
states. The events are required to contain at least one primary vertex with
reconstructed $z$ position within 24 cm  of the geometric center of the detector
and $xy$ position within 2 cm of the beam interaction region.

The W$\gamma \to \ell \nu \gamma $  final state is characterized by a
prompt, energetic, and isolated lepton, significant \MET due
to the presence of the neutrino from the W boson decay, and a prompt
isolated photon.
The basic event selection is similar for the electron and muon
channels: we require a charged lepton, electron or muon, with $\pt>20$~GeV,
which must satisfy the trigger requirements;
one photon with transverse energy $\ETG > 10 \gev$, and the \MET in the event
exceeding 25 GeV. As mentioned before, the photon must be separated from the
lepton by $\Delta R(\ell, \gamma)>0.7$.
For the e$\nu\gamma$ channel, the electron candidate must
satisfy the tight electron selection criteria. If the event has an additional
electron that satisfies the loose electron selection, we reject the event
to reduce contamination from Z/$\gamma^* \to \text{ee}$ processes.
For $\mu\nu\gamma$, we reject the event if a second muon is found
with $p_\text{T} > 10$~GeV.

After the full selection, 452 events are selected in the e$\nu\gamma$ channel
and 520 events are selected in the $\mu\nu\gamma$ channel. No events
have more than one photon candidate in the final state. The background
from misidentified jets estimated in data amounts to
$220 \pm 16~\text{(stat.)} \pm 14~\text{(syst.)}$
events for the e$\nu\gamma$ final state, and
$261 \pm 19~\text{(stat.)} \pm 16~\text{(syst.)}$ events
for the $\mu\nu\gamma$ final state. Backgrounds from other sources,
such as the Z$\gamma$ process in which one of the leptons from the Z boson
decay does not pass the reconstruction and identification criteria
and diboson processes where one of the electrons is misreconstructed as a
photon, are estimated from MC simulation and found to be
$7.7 \pm 0.5$ and $16.4 \pm 1.0$ for W$\gamma\to \text{e}\nu\gamma$ and
W$\gamma \to \mu\nu\gamma$, respectively. A larger contribution from Z$\gamma$
background in  the muon channel is due to a smaller pseudorapidity coverage for
muons, thus increasing the probability for one of the Z decay muons to be
lost, which results also in an overestimated value of the measured missing
energy in such events as the lost muon cannot be taken into account in the \MET
determination. The W$\gamma \to \tau\nu\gamma$ production,
with subsequent $\tau\to \ell\nu\nu$ decay, also contributes at the few percent
level to the $e\nu\gamma$ and $\mu\nu\gamma$ final states. We rely on MC
simulation to estimate this contribution.
The $E_T$ distribution for photon candidates in events passing the
full \Wg\ selection is given in Fig.~\ref{fig:ETG_wg}.

The three tree-level W$\gamma$ production processes interfere
with each other, resulting in a radiation-amplitude zero (RAZ) in
the angular distribution of the
photon~\cite{razTheory1, razTheory2, razTheory3, razTheory4, razTheory5}.
The first evidence for RAZ in \Wg\ production was observed by the
D0 collaboration~\cite{tevatron_wg2} using the charge-signed
rapidity difference $Q_\ell \times \Delta\eta$ between the photon candidate
and the charged lepton candidate from the $W$ boson decay~\cite{raz1}.
In the SM, the location of the dip minimum is located at
$Q_\ell \times \Delta \eta = 0$ for pp collisions.
Anomalous W$\gamma$ production can result in a flat distribution of the
charge-signed rapidity difference.

In Fig.~\ref{fig:raz} we plot the charge-signed rapidity difference in
background-subtracted data with an additional requirement on the transverse mass
of the photon, lepton, and \MET to exceed 90 GeV, to reduce the contribution from
FSR \Wg\ production. The agreement between background-subtracted
data and MC prediction is reasonable, with a Kolmogorov-Smirnov 
test~\cite{Kolmogorov33,Smirnov48} result of 57\%.

\begin{figure}[tb]
  \begin{center}
    \includegraphics[width=0.8\textwidth]{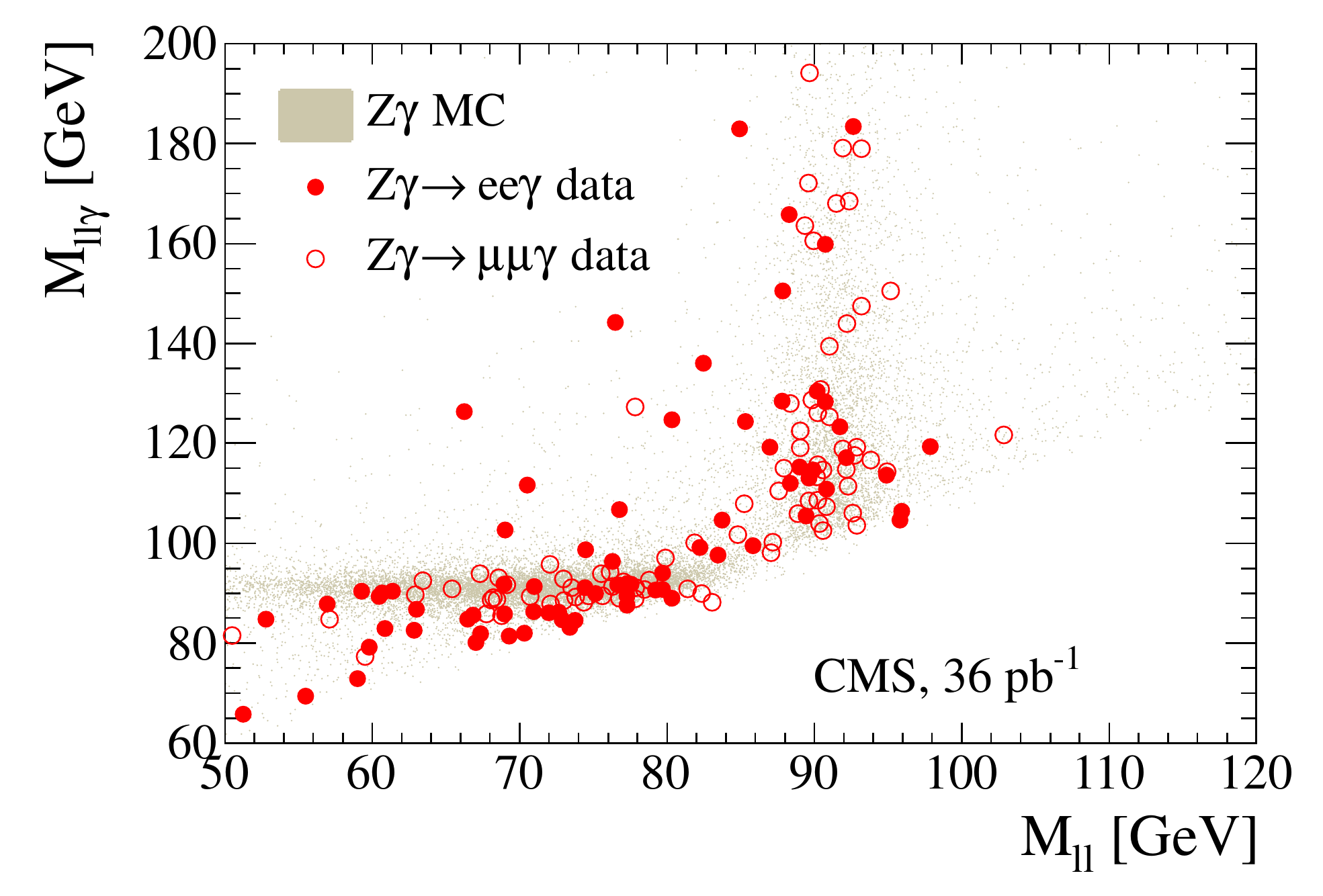}
    \caption{
Distribution of the $\ell\ell\gamma$ invariant mass as a function
of the dilepton invariant mass for selected Z$\gamma$ candidates in the
electron (filled circles) and muon (open circles) final states. The data
accumulation at $M_{\ell\ell\gamma} \simeq M_Z$ corresponds to FSR events, while
the data at $M_{\ell\ell} \simeq M_Z$ correspond to ISR events.
}
   \label{fig:Zg_banana}
  \end{center}
\end{figure}

Events in the Z$\gamma$ sample are selected by requiring a pair of
electrons or muons, each with transverse momentum $\pt>20$~GeV, forming
an invariant mass above 50 GeV.  One of these leptons must satisfy
the trigger requirements. The events are further required to have
a photon candidate passing the selection criteria with transverse energy
\ETG\ above 10 \gev. The photon must be separated from any of the two
charged leptons by $\Delta R(\ell, \gamma)>0.7$.
After applying these selection criteria we observe
81 events in the ee$\gamma$ final state and 90 events in the
$\mu\mu\gamma$ final state. No events are observed with more than one
photon candidate.
The Z+jets background to these final states is estimated to be
$20.5 \pm 1.7~\text{(stat.)} \pm 1.9~\text{(syst.)}$ and
$27.3 \pm 2.2~\text{(stat.)} \pm 2.3~\text{(syst.)}$, respectively.
Other backgrounds from multijet QCD, $\gamma+\text{jets}$, t$\bar{\text{t}}$,
and other diboson processes contribute less than one event in each of the two channels
and are therefore neglected in this analysis.
The $E_T$ distribution of the photon candidates in the selected  Z$\gamma$ candidate
events is shown in Fig.~\ref{fig:ETG_zg}. The distribution of
the $\ell \ell \gamma$ mass
as a function of the dilepton mass is displayed in Fig.~\ref{fig:Zg_banana}. We
observe good agreement between data and the SM prediction.

The measurement of the cross sections is based on the formula
\begin{equation}
\sigma = \frac{N_{\mathrm{data}} - N_{\mathrm{bkg}}}{A \epsilon \mathcal{L}},
\label{eq:xs}
\end{equation}
where $N_\mathrm{data}$ is the number of observed events, $N_\mathrm{bkg}$ is the
number of estimated background events, $A$ is the fiducial and kinematic
acceptance of the selection criteria, $\epsilon$ is the selection efficiency
for events within the acceptance, and  $\mathcal{L}$ is the integrated
luminosity. The acceptance is determined relative to the phase space defined by
the cuts $\ETG\ >  10~\gev$ and $\Delta R(\ell, \gamma)>0.7$, and in addition by
$M_{\ell\ell} > 50 \gev$ for \Zg.
We determine the product $A\cdot \epsilon$ from MC simulations
and apply correction factors $\rho$ to account for differences
in efficiencies between data and simulations.
These correction factors come from efficiency ratios $\rho = \epsilon/\epsilon_{\mathrm{sim}}$ derived by measuring
$\epsilon$ and $\epsilon_\mathrm{sim}$ in the same way on data and simulations, respectively, following the
procedure used in the inclusive W and Z measurement~\cite{wzxs}.

Systematic uncertainties are grouped into three categories.
In the first group, we combine the uncertainties that affect the product
of the acceptance, reconstruction, and identification efficiencies
of final state objects, as determined from Monte Carlo simulation.
These include uncertainties on lepton and photon energy scales and resolution,
effects from pile-up interactions, and uncertainties in the parton
distribution functions (PDFs). Lepton energy scale and resolution
effects are estimated
by studying the invariant mass of $\mathrm{Z}\to \ell\ell$ candidates, while the
photon energy scale and resolution uncertainty comes from ECAL calibration
studies which are further cross-checked with
the Z$\gamma$ FSR study.
The uncertainty due to the PDFs is estimated following Ref.~\cite{CTEQ}.
The second group includes the systematic uncertainties affecting
the data vs.\ simulation correction factors $\rho$ for the efficiencies
of the trigger, reconstruction, and identification requirements.
These include lepton trigger, lepton and photon reconstruction and
identification, and \MET efficiencies for the W$\gamma$ process.
The lepton efficiencies are determined by the ``tag-and-probe''
method~\cite{wzxs} in the same way for data and simulation,
and the uncertainty on the ratio of efficiencies is taken as a systematic
uncertainty. The third category comprises uncertainties on the
background yield. These are dominated by the uncertainties on the
data-driven W+jets and Z+jets background estimation.
These include systematic uncertainties due to the modeling of the
$E_\text{T}^\gamma$-dependent ratio and the uncertainty due to
the $\gamma+\text{jets}$ contribution.
Finally, an additional uncertainty due to the measurement of the integrated
luminosity is considered. This uncertainty is 4\%~\cite{lumi2}.
All systematic uncertainties for the W$\gamma$ and Z$\gamma$ channels are summarized
in Table~\ref{tab:systematics}.

\begin{table}[h]
  \begin{center}
   \caption{Summary of systematic uncertainties.}
    \label{tab:systematics}
    \begin{tabular}{l|c|c||c|c}
      \hline\hline
                                      & W$\gamma\to\ $e$\nu\gamma$ &  W$\gamma\to\mu\nu\gamma$ & Z$\gamma\to$ee$\gamma$ & Z$\gamma\to\mu\mu\gamma$   \\ \hline
      Source                                         & \multicolumn{4}{c}{Effect on $A \cdot \epsilon_{\mathrm{MC}}$}   \\ \hline
      Lepton energy scale                            & 2.3\%            & 1.0\%           & 2.8\%      & 1.5\% \\
      Lepton energy resolution                       & 0.3\%            & 0.2\%           & 0.5\%      & 0.4\% \\
      Photon energy scale                            & 4.5\%            & 4.2 \%          & 3.7\%      & 3.0\% \\
      Photon energy resolution                       & 0.4\%            & 0.7\%           & 1.7\%      & 1.4\% \\
      Pile-up                                        & 2.7\%            & 2.3\%           & 2.3\%      & 1.8\% \\
      PDFs                                           & 2.0\%            & 2.0\%           & 2.0\%      & 2.0\% \\ \hline
      Total uncertainty on $A \cdot \epsilon_\mathrm{MC}$   & 6.1\%            & 5.2\%           & 5.8\%      & 4.3\% \\ \hline\hline
                                                     & \multicolumn{4}{c}{Effect on $\epsilon_{\mathrm{data}}/\epsilon_{\mathrm{MC}}$} \\ \hline
      Trigger				             & 0.1\% 	        & 0.5\%           & $<0.1\%$   & $<0.1\%$ \\
      Lepton identification and isolation            & 0.8\% 	 	& 0.3\%           & 1.1\%      & 1.0\% \\
      \MET selection                                 & 0.7\%            & 1.0\%           & N/A        & N/A   \\
      Photon identification and isolation            & 1.2\%            & 1.5\%           & 1.0\%      & 1.0\% \\ \hline
      Total uncertainty on $\epsilon_\mathrm{data}/\epsilon_\mathrm{MC}$
                                                     & 1.6\%            & 1.9\%           & 1.6\%      & 1.5\% \\ \hline\hline
      Background                                     & 6.3\%            & 6.4\%           & 9.3\%      & 11.4\% \\  \hline
      Luminosity  		                     & \multicolumn{4}{c}{ 4\%} \\ \hline
      \hline
    \end{tabular}
  \end{center}
\end{table}

We find the cross section for \Wg\ production for $\ETG >10$~GeV and
$\DR(\ell,\gamma)>0.7$ to be
$\sigma(\mathrm{pp}\to \Wg +X) \times \mathcal{B}(\text{W}\to \text{e}\nu) =
        57.1  \pm 6.9 \text{~(stat.)}
         \pm 5.1 \text{~(syst.)}
        \pm 2.3 \text{~(lumi.)}$~pb and
$\sigma(\mathrm{pp}\to \Wg +X) \times \mathcal{B}(\text{W}\to\mu\nu) =
        55.4 \pm 7.2\text{~(stat.)} \pm 5.0 \text{~(syst.)}
        \pm 2.2 \text{~(lumi.)}$~pb.
Taking into account correlated uncertainties between these two results,
due to photon identification,
energy scale, resolution, data-driven background, and signal modeling,
and following the Best Linear Unbiased Estimator method~\cite{blue}, we
measure the combined cross section to be
$\sigma(\mathrm{pp}\to\Wg+X) \times \mathcal{B}(\text{W}\to \ell\nu) =
56.3 \pm 5.0~\text{(stat.)} \pm 5.0~\text{(syst.)} \pm 2.3~\text{(lumi.)}$~pb.
This result agrees well with the NLO prediction~\cite{baurWg}
of $49.4\pm3.8$~pb.

The Z$\gamma$ cross section within the requirements $\ET^\gamma>10$~GeV,
$\DR(\ell,\gamma)>0.7$, and $m_{\ell\ell}>50$~GeV, is measured to be
$\sigma(\mathrm{pp}\to\text{Z}\gamma+X) \times \mathcal{B}(\text{Z} \to \text{ee}) =
9.5 \pm 1.4~(\text{stat.}) \pm 0.7~(\text{syst.}) \pm 0.4~(\text{lumi.})$~pb
for the ee$\gamma$ final state, and
$\sigma(\mathrm{pp}\to\text{Z}\gamma+X) \times \mathcal{B}(\text{Z}\to \mu\mu) =
9.2 \pm 1.4~(\text{stat.}) \pm 0.6~(\text{syst.}) \pm 0.4~(\text{lumi.})$~pb
for the $\mu\mu\gamma$ final state. The combination of the two results yields
$\sigma(\mathrm{pp}\to\text{Z}\gamma+X) \times \mathcal{B}(\text{Z}\to \ell\ell) =
9.4 \pm 1.0~(\text{stat.}) \pm 0.6~(\text{syst.}) \pm 0.4~(\text{lumi.})$~pb.
The theoretical NLO prediction~\cite{baur} is $9.6\pm0.4$~pb, which is in
agreement with the measured value.

Given the good agreement of both the measured cross sections
and the \ETG\ distributions with the corresponding SM predictions,
we proceed to set limits on anomalous TGCs.
The most general Lorentz-invariant Lagrangian that describes the
WW$\gamma$ coupling has seven independent dimensionless
couplings $g_1^\gamma$, $\kappa_\gamma$, $\lambda_\gamma$,
$g^\gamma_4$, $g_5^\gamma$, $\tilde{\kappa}_\gamma$, and
$\tilde{\lambda}_\gamma$~\cite{hagiwara1}.
By requiring $CP$ invariance and $SU(2) \times U(1)$ gauge invariance only two
independent parameters remain: $\kappa_\gamma$ and $\lambda_\gamma$.
In the SM, $\kappa_\gamma = 1$ and $\lambda_\gamma = 0$. We define aTGCs to be
deviations from the SM predictions, so instead of using $\kappa_\gamma$
we define $\Delta \kappa_\gamma \equiv \kappa_\gamma - 1$.
While these couplings have no physical meaning as such, they are related
to the electromagnetic moments of the W boson,
\begin{equation}
 \mu_\text{W} = \frac{e}{2M_\text{W}}(2 + \Delta\kappa_\gamma + \lambda_\gamma), \quad
 Q_\text{W}   = - \frac{e}{M_\text{W}^2} ( 1 + \Delta\kappa_\gamma - \lambda_\gamma),
\end{equation}
where $\mu_\text{W}$ and $Q_\text{W}$ are the magnetic dipole and electric
quadrupole moments of the W boson, respectively.

For the ZZ$\gamma$ or Z$\gamma\gamma$ couplings,
the most general Lorentz-invariant and gauge-invariant vertex is described
by only four parameters $h_i^V$  ($i=1,2,3,4$; $V=\gamma,\text{Z}$)~\cite{baur}.
By requiring $CP$ invariance, only two parameters, $h_3^V$ and $h_4^V$,
remain. The SM predicts these couplings to vanish at tree level.
Simulated samples of W$\gamma$ and Z$\gamma$ signals for a grid of
aTGCs values are produced similarly to the SM signal W$\gamma$ and Z$\gamma$
samples described above. A grid of $\lambda_\gamma$ and $\Delta\kappa_\gamma$
values is used for the WW$\gamma$ coupling, and a grid of $h_3$ and $h_4$
values is used for the ZZ$\gamma$ and Z$\gamma\gamma$ couplings.

\begin{figure}[tb]
\begin{center}
\includegraphics[width=0.49\textwidth]{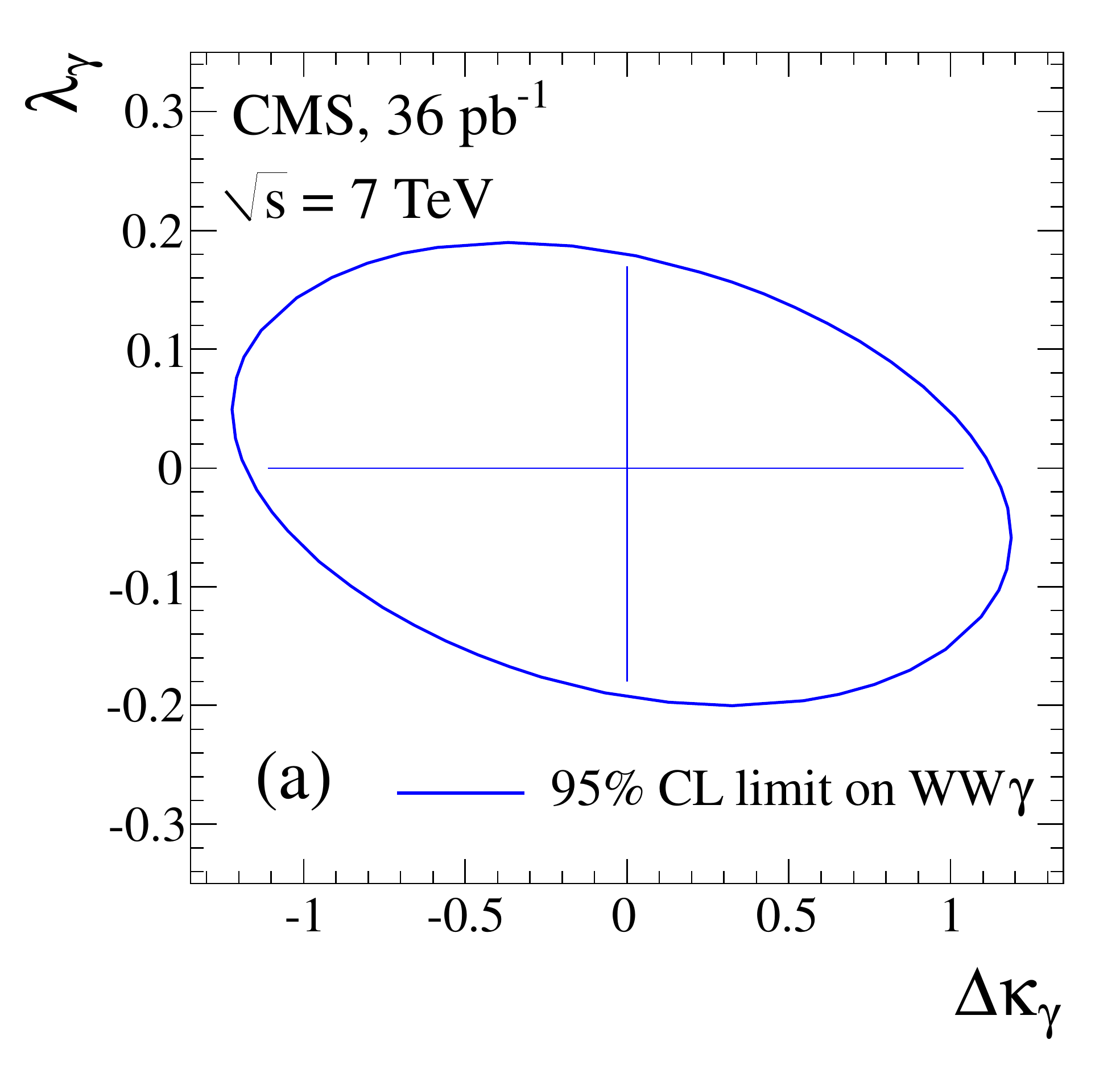}
\includegraphics[width=0.49\textwidth]{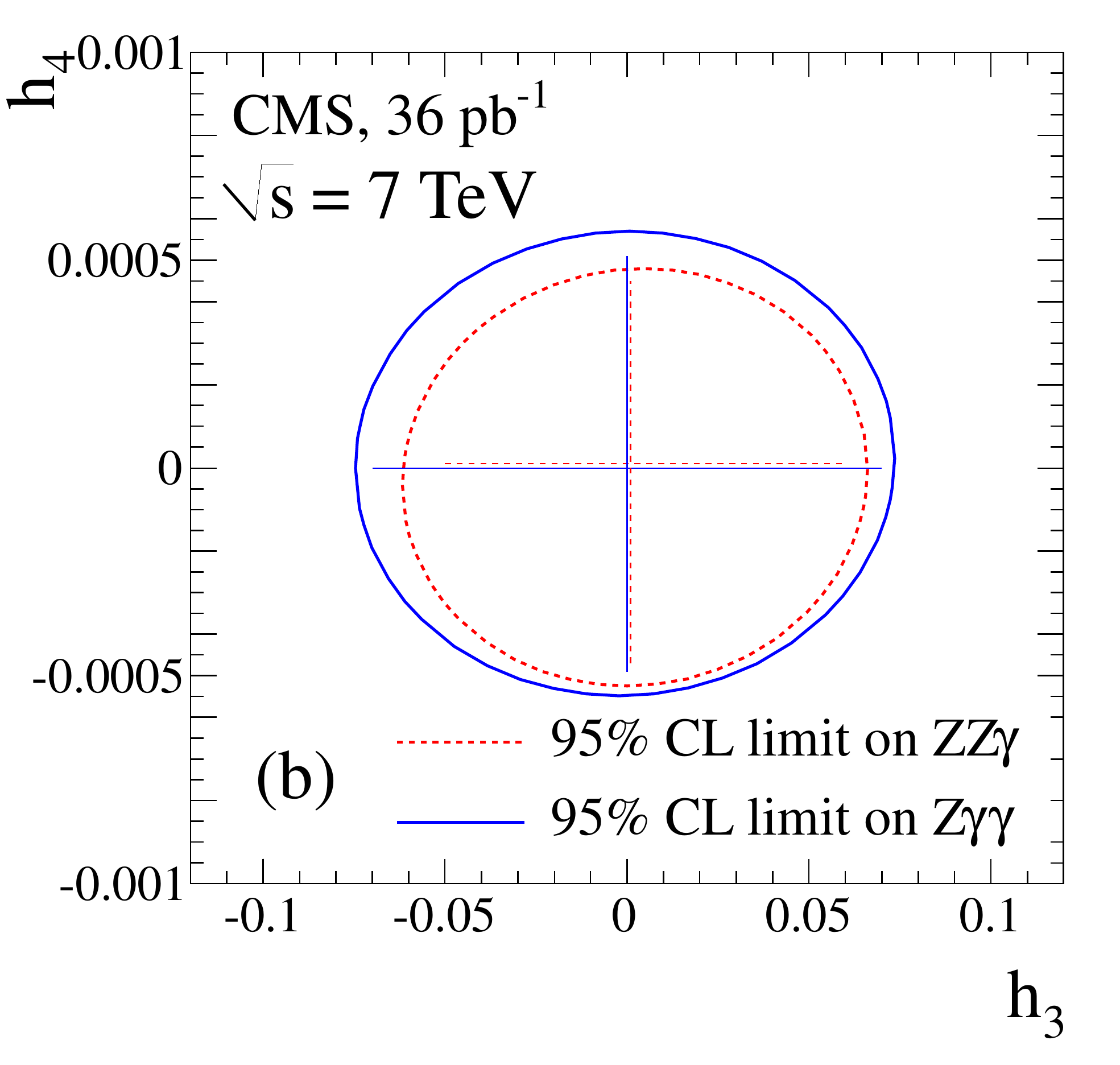}
\caption{Two-dimensional 95\% CL limit contours (a) for the WW$\gamma$ vertex couplings
$\lambda_\gamma$ and $\Delta\kappa_\gamma$ (blue line), and (b) for the
ZZ$\gamma$ (red dashed line) and Z$\gamma\gamma$ (blue solid line) vertex
couplings $h_3$ and $h_4$ assuming no energy dependence on the couplings.
One-dimensional 95\% CL limits on individual couplings are given as
solid lines.}
   \label{fig:atgc_contour}
  \end{center}
\end{figure}

Assuming Poisson statistics and log-normal distributions for the
generated samples and background systematic uncertainties
we calculate the likelihood of the observed photon $E_\text{T}$ spectrum
in data given the sum of the background and aTGCs \ETG~predictions for
each point in the grid of aTGCs values.
To extract the limits we parameterize the
expected yields as a quadratic function of the anomalous couplings.
We then form the probability of observing the number of events seen
in data in a given bin of the photon transverse energy using a Poisson
distribution with the mean given by the expected signal plus a data driven
background estimate and allowing for variations within the systematic
uncertainties. The confidence intervals are found using {\sc MINUIT}, profiling the
likelihood with respect to all systematic variations~\cite{stat}.
The resultant two-dimensional 95\% confidence level (CL) limits
are given in Fig.~\ref{fig:atgc_contour}.  To set one-dimensional 95\% CL
limits on a given anomalous coupling we set the other aTGCs to their respective
SM predictions. The results are summarized in Table~\ref{tab:atgc1DLimits}.

\begin{table}
\begin{center}
\caption{One dimensional 95\% CL limits on WW$\gamma$, ZZ$\gamma$, and Z$\gamma\gamma$ aTGCs.
}
\label{tab:atgc1DLimits}
\begin{tabular}{c|c|c} \hline \hline
WW$\gamma$                            & ZZ$\gamma$               & Z$\gamma\gamma$          \\ \hline
$-1.11 < \Delta\kappa_\gamma < 1.04$  & $-0.05   < h_3 < 0.06$   & $-0.07   < h_3 < 0.07$   \\
$-0.18 < \lambda_\gamma      < 0.17$  & $-0.0005 < h_4 < 0.0005$ & $-0.0005 < h_4 < 0.0006$ \\ \hline \hline
\end{tabular}
\end{center}
\end{table}

\begin{figure}[tb]
\begin{center}
\includegraphics[width=0.99\textwidth]{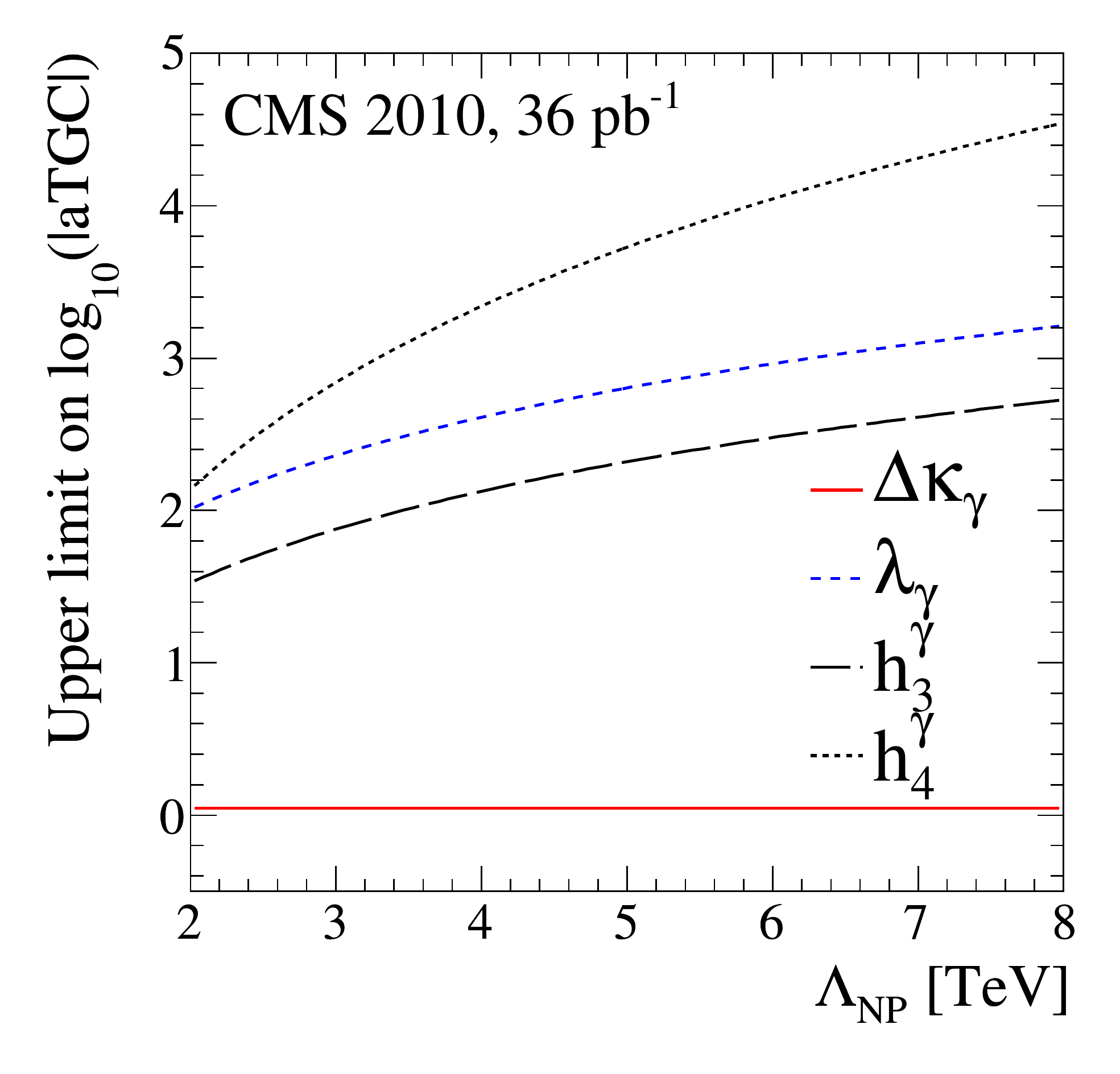}
\caption{Upper 95\% CL limits on $\log_{10}(|\text{aTGC}|)$ as a function
of $\Lambda_\text{NP}$ for $\Delta\kappa_\gamma$, $\lambda_\gamma$, $h_3^\gamma$, and
$h_4^\gamma$. Limits on the latter two couplings are similar to those for
$h_3^\text{Z}$ and $h_4^\text{Z}$. These limits refer to the formulation
in which the new physics Lagrangian terms are scaled with
$\alpha/\Lambda_\text{NP}^n$, where $\Lambda_\text{NP}$ is the characteristic
energy scale of new physics and $\alpha$ is the aTGC.
}
\label{fig:LambdaNP}
\end{center}
\end{figure}

All the non-SM terms in the effective Lagrangian are scaled with 
$\alpha/m_\text{V}^n$, where $\alpha$ is an aTGC,
$m_\text{V}$ is the mass of the gauge boson (W boson for the WW$\gamma$ coupling
and Z boson for ZZ$\gamma$ and Z$\gamma\gamma$ couplings), and $n$ is a power that is chosen to
make the aTGC dimensionless. The values of $n$ for $\Delta\kappa_\gamma$,
$\lambda_\gamma$, $h_3$, and $h_4$ are 0, 2, 2, and 4, respectively.
An alternative way to scale those new physics Lagrangian terms is with
$\alpha/\Lambda_\text{NP}^n$, where $\Lambda_\text{NP}$
is the characteristic energy scale of new physics~\cite{Mery:1987et}.
We present upper limits on aTGCs  for $\Lambda_\text{NP}$ values between 2 and 8 TeV
in Fig.~\ref{fig:LambdaNP}.

In summary, we have presented the first measurement of the W$\gamma$ and Z$\gamma$
cross sections in pp collisions
at $\sqrt{s} = 7$~TeV for $\ETG > 10$~GeV, $\Delta R(\gamma, \ell) > 0.7$, and
for the additional requirement on the dilepton invariant mass to exceed 50 GeV for the
\Zg\ process. We measured the W$\gamma$ cross section times the branching
fraction for the leptonic W decay to be
$\sigma(\mathrm{pp}\to\text{W}\gamma + X) \times \mathcal{B}(\text{W}\to \ell\nu) =
56.3 \pm 5.0~\text{(stat.)} \pm 5.0~\text{(syst.)} \pm 2.3~(\text{lumi.})$~pb.
This result is in good agreement with the NLO prediction of $49.4\pm3.8$~pb, where
the uncertainty includes both PDF and $k$-factor uncertainties.
The Z$\gamma$ cross section times the branching fraction for the leptonic Z decay was
measured to be
$\sigma(\mathrm{pp}\to\text{Z}\gamma + X) \times \mathcal{B}(\text{Z}\to \ell\ell) =
9.4 \pm 1.0~(\text{stat.}) \pm 0.6~(\text{syst.}) \pm 0.4~(\text{lumi.})$~pb,
which also agrees well with the NLO predicted value~\cite{baur}
of $9.6\pm0.4$~pb.
We also searched and found no evidence for anomalous WW$\gamma$,
ZZ$\gamma$, and Z$\gamma\gamma$ trilinear gauge couplings. We set the
first 95\% CL limits on these couplings at $\sqrt{s} = 7$~TeV.
These limits extend the previous results~\cite{lep,l3_1,l3_2,opal,tevatron_wg1, tevatron_wg2, tevatron_wg3, tevatron_zg0, tevatron_zg1, tevatron_zg2} on vector boson self-interactions at lower energies.

We wish to congratulate our colleagues in the CERN accelerator departments for the excellent
performance of the LHC machine. We thank the technical and administrative staff at CERN and
other CMS institutes, and acknowledge support from: FMSR (Austria); FNRS and FWO (Belgium);
CNPq, CAPES, FAPERJ, and FAPESP (Brazil); MES (Bulgaria); CERN; CAS, MoST, and NSFC (China);
COLCIENCIAS (Colombia); MSES (Croatia); RPF (Cyprus); Academy of Sciences and NICPB (Estonia);
Academy of Finland, ME, and HIP (Finland); CEA and CNRS/IN2P3 (France); BMBF, DFG, and HGF (Germany);
GSRT (Greece); OTKA and NKTH (Hungary); DAE and DST (India); IPM (Iran); SFI (Ireland); INFN (Italy);
NRF and WCU (Korea); LAS (Lithuania); CINVESTAV, CONACYT, SEP, and UASLP-FAI (Mexico); PAEC (Pakistan);
SCSR (Poland); FCT (Portugal); JINR (Armenia, Belarus, Georgia, Ukraine, Uzbekistan); MST and MAE (Russia);
MSTD (Serbia); MICINN and CPAN (Spain); Swiss Funding Agencies (Switzerland); NSC (Taipei); TUBITAK and
TAEK (Turkey); STFC (United Kingdom); DOE and NSF (USA).

\bibliography{auto_generated}   

\clearpage
\clearpage

\cleardoublepage \appendix\section{The CMS Collaboration \label{app:collab}}\begin{sloppypar}\hyphenpenalty=5000\widowpenalty=500\clubpenalty=5000\textbf{Yerevan Physics Institute,  Yerevan,  Armenia}\\*[0pt]
S.~Chatrchyan, V.~Khachatryan, A.M.~Sirunyan, A.~Tumasyan
\vskip\cmsinstskip
\textbf{Institut f\"{u}r Hochenergiephysik der OeAW,  Wien,  Austria}\\*[0pt]
W.~Adam, T.~Bergauer, M.~Dragicevic, J.~Er\"{o}, C.~Fabjan, M.~Friedl, R.~Fr\"{u}hwirth, V.M.~Ghete, J.~Hammer\cmsAuthorMark{1}, S.~H\"{a}nsel, M.~Hoch, N.~H\"{o}rmann, J.~Hrubec, M.~Jeitler, G.~Kasieczka, W.~Kiesenhofer, M.~Krammer, D.~Liko, I.~Mikulec, M.~Pernicka, H.~Rohringer, R.~Sch\"{o}fbeck, J.~Strauss, F.~Teischinger, P.~Wagner, W.~Waltenberger, G.~Walzel, E.~Widl, C.-E.~Wulz
% ...
\vskip\cmsinstskip
% ...
1:~~Also at CERN, European Organization for Nuclear Research, Geneva, Switzerland\\
% ...

\end{sloppypar}
\end{document}